\documentclass[fleqn,usenatbib]{mnras}

\usepackage{newtxtext,newtxmath}

\usepackage[T1]{fontenc}
\usepackage[flushleft]{threeparttable}

\newcommand\chandra{{\it Chandra}}

\newcommand\kms{\ifmmode {\rm~km\ s}^{-1} \else ~km s$^{-1}$\fi}
\newcommand\Hunit{\ifmmode {\rm~km\ s}^{-1}\ {\rm Mpc}^{-1}
        \else ~km s$^{-1}$ Mpc$^{-1}$\fi}
\newcommand\ctssec{\ifmmode {\rm~count\ s}^{-1} \else ~count s$^{-1}$\fi}
\newcommand\ergsec{\ifmmode {\rm~erg\ s}^{-1} \else
        ~erg s$^{-1}$\fi}
\newcommand\funit{\ifmmode {\rm~erg\ s}^{-1}\;{\rm cm}^{-2} \else
        ~ergs s$^{-1}$ cm$^{-2}$\fi}
\newcommand\phflux{\ifmmode {\rm~photon\ s}^{-1}\;{\rm cm}^{-2}
        \else   ~photon s$^{-1}$ cm$^{-2}$\fi}
\newcommand\efluxA{\ifmmode {\rm~erg\ s}^{-1}\;{\rm cm}^{-2}\;{\rm
        \AA}^{-1} \else ~erg s$^{-1}$ cm$^{-2}$ \AA$^{-1}$\fi}
\newcommand\efluxHz{\ifmmode {\rm~erg\ s}^{-1}\;{\rm cm}^{-2}\;{\rm
        Hz}^{-1} \else ~erg s$^{-1}$ cm$^{-2}$ Hz$^{-1}$\fi}
\newcommand\cc{\ifmmode {\rm~cm}^{-3} \else cm$^{-3}$\fi}
\newcommand\FWHM{\ifmmode {\rm~FWHM} \else ${\rm~FWHM}$\fi}
\newcommand\Lsun{\ifmmode L_{\odot} \else $L_{\odot}$\fi}

\newcommand\hbeta{\ifmmode {\rm H}\beta \else H$\beta$\fi}
\newcommand\Kalpha{\ifmmode {\rm K}\alpha \else K$\alpha$\fi}
\newcommand\nh{\ifmmode N_{\rm H} \else N$_{\rm H}$\fi}

\newcommand{\Msun}{\ensuremath{\rm M_{\odot}}}
\newcommand\Zsun{\ifmmode Z_{\odot} \else $Z_{\odot}$\fi}

\usepackage{xcolor}

\DeclareRobustCommand{\VAN}[3]{#2}
\let\VANthebibliography\thebibliography
\def\thebibliography{\DeclareRobustCommand{\VAN}[3]{##3}\VANthebibliography}

\usepackage{graphicx}	
\usepackage{amsmath}	
\usepackage{gensymb}
\usepackage{soul}

\title[Radio mini-halo in Abell~795]{Sloshing and spiral structures breeding a putative radio mini-halo in the environment of a cool-core cluster Abell~795}

\author[Kadam et al.]{S. K. Kadam$^{1}$, Sameer Salunkhe$^{2,3}$, N. D. Vagshette$^{4}$, Surajit Paul$^{5,6}$$\thanks{\color{blue}{surajit.paul@manipal.edu}}$, S. S. Sonkamble$^{7}$, P. K. Pawar$^{8}$ \newauthor and M. K. Patil$^{1}$\thanks{\color{blue}{patil@associates.iucaa.in}} \\ \\
$^{1}$School of Physical Sciences, Swami Ramanand Teerth Marathwada University, Nanded-431 606, India\\
$^{2}$National Centre for Radio Astrophysics (NCRA), Tata Institute of Fundamental Research (TIFR), Pune 411007, India\\
$^{3}$Department of Physics, Savitribai Phule Pune University, Pune 411007, India\\
$^{4}$Department of Physics and Electronics, Maharashtra Udayagiri Mahavidyalaya, Udgir, Dist. Latur-413517, India\\
$^{5}$Manipal Centre for Natural Sciences, Centre of Excellence, Manipal Academy of Higher Education, Manipal 576104, India \\
$^{6}$Raman Research Institute, Raman Research Institute, Bangalore, 560080, India\\
$^{7}$Centre for Space Research, North-West University, Potchefstroom 2520, South Africa\\
$^{8}$ Department of Physics, Deogiri College, Aurangabad - 431005, India\\
}

\begin{document}
\pagerange{\pageref{firstpage}--\pageref{lastpage}} \pubyear{2023}
\maketitle
\label{firstpage}
\begin{abstract}
Spiral structures and cold fronts in X-rays are frequently observed in cool core galaxy clusters. However, studies on radio mini-haloes associated with such spirals and their physical connections are rare. Here, we present the detection of an extended diffuse radio emission entrained in the X-ray spiral structure in a known cool core cluster Abell~795 (A795). Though the cool core is a sign of the relaxed nature of the clusters, our re-analysed 30~ks \textit{Chandra} X-ray data of cluster A795 confirms the presence of an interesting log spiral structure of X-ray deficit region complemented by an X-ray excess counter spiral in the residual map, exposing its dynamical activity. Our new analysis of 150~\&~325~MHz GMRT archival data of the cluster confirms the detection of a $\sim180$~kpc ultra-steep ($\alpha\sim-2.7$) diffuse radio structure which was previously reported as a candidate radio mini halo from low sensitive survey maps. This radio emission spans the entire spiral structure ($\sim186$~kpc), enclosed by two previously reported cold fronts. Furthermore, SDSS~DR13 optical spectra, as well as GALEX’s FUV data, show a considerably low total star formation rate of 2.52~M$_{\odot}$~yr$^{-1}$ and having no significant variation in metallicity distribution. We argued that the two-phase (hot and cold) plasma at the cluster core with differential velocity has probably caused the spiral formation and has redistributed the secondary electrons from the central BCG or the pre-accelerated electrons which have been (re-)accelerated by the sloshing turbulence to form the observed candidate radio mini-halo structure. This has been supported by a few previous studies that indicate spiral formation and sloshing turbulence may quench star formation and facilitate smooth metallicity distribution by mixing the gas in the core.
\end{abstract}

\begin{keywords}
galaxies: clusters: individual: Abell~795, X-rays: galaxies: clusters, galaxies: clusters: intracluster medium, radio continuum: general
\end{keywords}

\section[1]{Introduction}
Galaxy clusters are the gravitationally bound most massive aggregates in the Universe. Their formation through accretion and mergers of smaller subclusters and groups act as the direct probe to the hierarchical structure formation \citep{1980lssu.book.....P}.
The knowledge of the dynamical states of a cluster is not only important for understanding its physical properties, but it also sheds light on the cosmological properties \citep{2003MNRAS.341.1311S}. In the last few decades, the dynamical states of hundreds of galaxy clusters have been extensively investigated by studying their substructures using the X-ray images and spectroscopic data \citep{2005AJ....129.1350S, 2005ApJ...628..655V, 2008MNRAS.383..879A, 2010A&ARv..18..127B, 2008A&A...483...35S}. Based on the dynamical states, they are broadly classified into two classes, the relaxed and the unrelaxed or disturbed clusters. Relaxed clusters are particularly influenced by the feedback of Active Galactic Nuclei (AGN) to their intracluster medium  \citep[e.g.][]{2012ARA&A..50..455F}.

High angular resolution imaging of the cores of galaxy clusters with \textit{Chandra} X-ray telescope has unravelled the imprint of the interaction of radio jets from the central AGN with the surrounding ICM in the form of X-ray deficit bubbles (cavities), shock fronts \citep{2003MNRAS.344L..43F, 2005ApJ...635..894F, 2006MNRAS.366..417F, 2007ARA&A..45..117M, 2009AIPC.1201..233G, 2011ApJ...726...86R, 2017MNRAS.466.2054V}. Some of the clusters also exhibit spectacular features such as spiral patterns, filamentary structures etc. In this context, a few well-studied systems are Abell 2142, Bullet cluster, Perseus, NGC 5813, Abell 2390, Zwcl 2701 \citep{2000ApJ...541..542M,2002ApJ...567L..27M,2003ApJ...590..225C,2011ApJ...726...86R,2015Ap&SS.359...61S,2016MNRAS.461.1885V}. It is well established that the relaxed clusters in general host cool cores at their centre \citep{1992MNRAS.258..177E, 2001MNRAS.328L..37A, 2005MNRAS.359.1481B, 2005ApJ...628..655V, 2008MNRAS.383..879A, 2013Ap&SS.345..183P, 2016MNRAS.461.1885V, 2015Ap&SS.359...61S,2019MNRAS.484.4113K}, while, the unrelaxed or mildly disturbed clusters might have cool cores with sloshing spirals in 70\% of cases which are generally formed by mergers \citep[e.g.][]{1999ApJ...511...65J,2012MNRAS.420.2120M}.

The formation of spiral structures in the cluster environments is the ramification of the gas sloshing due to bulk motions of cold gas displaced in the cores of galaxy clusters. As discussed in \citet{2006ApJ...650..102A,2007PhR...443....1M,2010ApJ...717..908Z} such phenomena may further be attributed to the passage of substructures (mass clumps) that induce gravitational disturbances in the ICM. This proposition is also supported by the numerical simulations by several researchers \citep[e.g.,][]{2006ApJ...650..102A,2013A&A...556A..44R}. These simulations have demonstrated that the steep central entropy may be responsible for initiating the sloshing which probably have been observed in the ICM of several galaxy clusters such as Perseus \citep{2003ApJ...590..225C}, Abell 85 \citep{2019MNRAS.483.1744I}, Abell 2029 \citep{2013ApJ...773..114P}, Abell 1644 \citep{2010ApJ...710.1776J}, Abell 2052 \citep{2003ApJ...585..227B}, Abell 2142 \citep{2018ApJ...863..102L}, Abell 1763 \citep{2018ApJ...868..121D}. Simulations of two (hot and cold) phase plasma with differential flow velocity also found to produce spiral structures in the cluster centre \citep{Keshet_2012}. In an alternative scenario of transverse cooling flow, the gas trapped in the cooling region or inside the BCG potential, after hitting the boundary, reverses and starts in-falling toward the BCG. The rotational motion to this infalling gas is set due to the action of Coriolis force in BCG frame forming spiral structures \citep{Inoue_2022PASJ}. Notably, roughly half of the sample of clusters studied by \cite{2010A&A...511A..15L} exhibit spiral-like structures.

Transient dynamical effects in galaxy clusters, such as mergers, have their manifestation in non-thermal features such as cluster radio emissions \citep{brunetti_2014IJMPD..2330007B, van_Weeren_2019SSRv..215...16V,2023JApA...44...38P}. However, the clusters in their relaxed state associated with the cool cores too, occasionally harbour radio mini-haloes around their central BCG \citep{2004NewAR..48.1137F, 2012A&ARv..20...54F}. These are steep spectrum ($\alpha \lesssim -1$, where $\alpha$ is spectral index relating the radio flux density $S$ and observed frequency $\nu$ by a relation $S\propto \nu^\alpha$), diffuse radio emission of size ranging 100-500 kpc and are almost co-spatial with the cool-core of the host cluster. Their origin is usually discussed under two probable frameworks. The one that invokes turbulent re-acceleration, where the turbulence in the ICM is generated either due to a minor and/or off-axis merger keeping the core intact or by the multiphase gas flows of the ICM on the central galaxy. The seed electrons for this mechanism are mildly relativistic pre-existing cosmic-ray electrons, possibly provided by the central BCG itself \citep{gitti_2002A&A...386..456G, zuhone_2013ApJ...762...78Z}. The other one is the hadronic scenario, where the cosmic ray electrons are continuously injected by collisions between the cosmic ray protons and the thermal proton population in the ICM and may travel over larger distances due to their lower emissivity \citep{pformmer_2004A&A...413...17P}. The former scenario assumes a balance between the injected turbulence with the induced gas sloshing motions, leaving intact the overall relaxed morphology of the parent cluster. It is also possible that the secondary electrons may get re-accelerated by the turbulence to produce the mini-halo type radio emission - described as the hybrid model by \citet{riseley_2022MNRAS.512.4210R,riseley_2023MNRAS.524.6052R}

Radio studies of clusters with sloshing have revealed the mini-haloes, often surrounded by cold fronts \citep{2001ApJ...555..205M, 2008ApJ...675L...9M, 2010A&A...516A..32G, 2013A&A...556A..44R}. Even though, cool-cores are a sign of a relaxed cluster environment, the presence of mini-haloes or spiral structures indicates dynamical activities. Thus, studying cool core clusters with mini-haloes and spiral features by acquiring multi-wavelength information is the key to understanding this contrasting dynamical situation in some galaxy clusters.

We find Abell 795 an ideal example of such a system which hosts a cool core and a pair of cold fronts associated with the sloshing activity as reported by \cite{2021MNRAS.503.4627U} in a recent study. The cool core is found to be weak with a mass deposition rate of $\leq$ 14 \Msun $yr^{-1}$ and the presence of disturbances. The cluster has been classified as morphologically disturbed \citep{2015A&A...575A..48S} and having unrelaxed ICM \citep{2016ApJ...823..116Z, 2015MNRAS.449..199M}. The signatures of dynamical disturbances in the ICM are evident in the form of a pair of X-ray depressions or cavities within 30 kpc and a pair of cold fronts at $\sim$ 50 kpc and $\sim$ 178 kpc on either sides of the central BCG peak \citep{2021MNRAS.503.4627U}. They have also discovered a spiral structure of the cool ICM possibly arising due to sloshing and is held responsible for the formation of the surface brightness discontinuities. The presence of the coldest gas distinctly aligns with this spiral structure of the ICM which is seen in the temperature map presented in \citet{2021AN....342.1207U}. Spectral analysis of the ICM in this cluster has also enabled them to measure the temperature jumps at these discontinuities implying that they are indeed the cold fronts. Using radio data from the VLA and TGSS, the authors have also reported the existence of a candidate radio mini-halo. Furthermore, they have shown the balance between the radio power and the kinetic power injected into the ICM by the central AGN. However, the low fidelity of these radio images restricted them from mapping the radio substructures, and thus providing any information on the nature and origin of the radio emission. 

Although the dynamical state and cooling properties of the intra-cluster medium of A795 have been reported by \citet{2021MNRAS.503.4627U}, detailed analysis of the observed spiral structure, especially thermal non-thermal connection using deeper radio maps were lacking in their study, which we compensate by analyzing 325 and 150 MHz GMRT archival data.  
Furthermore, we revisit the results from {\it{Chandra}} X-ray maps published in \citet{2021MNRAS.503.4627U} by reanalysing the same data, especially the features related to the spiral structure. We also studied the effect of the spiral motion on the star-forming activity in this cluster analysing optical spectra from SDSS data.

The paper is organized as follows: Section~\ref{red} provides a brief description on the data preparation and reduction. The results derived from the X-ray imaging, spectroscopy and radio data analysis are presented in Section~\ref{results}. Discussion based on the results derived from this study is presented in Section~\ref{disc}, while in Section~\ref{con} we summarize our findings.

We adopt the cosmological parameters as H$_0$= 73 km s$^{-1}$ Mpc$^{- 1}$, $\Omega_M$ = 0.27, $\Omega_{\Lambda}$ = 0.73.  Here, one arcsec at the redshift of $z=0.1359$ corresponds to 2.4 kpc on the physical scale of the images presented in this paper. 
\begin{figure}
\includegraphics[scale=0.45]{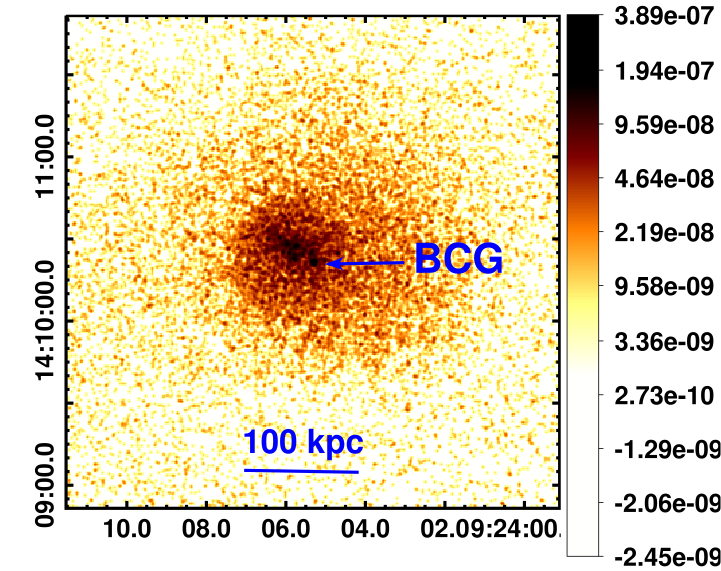}
\caption{Exposure corrected, point source removed, background subtracted 0.5 - 3 keV 3$\arcmin \times$ 3$\arcmin$ \textit{Chandra} X-ray image of Abell~795. For better visualization, this image has been smoothed with 2$\sigma$ Gaussian kernel.}
\label{raw}
\end{figure}

\section[2]{Observations and the Data reduction}
\label{red}
\subsection{X-ray Data}
Abell~795 was observed by the \chandra\, ACIS-S detector on 13 January 2010 (ObsID 11734, PI: Helen Russell) in `VFAINT' mode with the source focused on the back-illuminated Advanced CCD Imaging Spectrometer ACIS-S3 chip for a total exposure time of $\sim\,$ 30 ks. Level-1 event files corresponding to this observation ID were reprocessed using the \textit{chandra$\_$repro} within \textit{CIAO 4.9} and \textit{CALDB V 4.6.7}. Periods of high background flares over the entire chip were identified using the 3$\sigma$ clipping of the light curve task \textit{lc\_sigma\_clean} within CIAO, which yielded the net exposure time equal to 29.5 ks. Further, we created the background file by reprojecting the \textit{blanksky} frame corresponding to the ObsID 11734 and scaled it so as to match the hard X-ray photon count rates in the energy range (9-12 keV) with those in the source file. By utilizing \textit{blanksky$\_$image}, we generated an image that is background-subtracted and exposure-corrected. We then ran the \textit{wavdetect} tool of \textit{CIAO} to identify and remove the point sources, excluding the central brightest point source, within the science frame. The holes due to the removal of the point sources were filled in with local Poisson using \textit{dmfilth}. The resultant exposure corrected, point source removed, and background subtracted science image of A795 in the energy range 0.5 - 3 keV is shown in Fig.~\ref{raw}. For better representation, this image has been smoothed with a 2$\sigma$ Gaussian kernel.

\begin{figure*}
{
\includegraphics[scale=0.45]{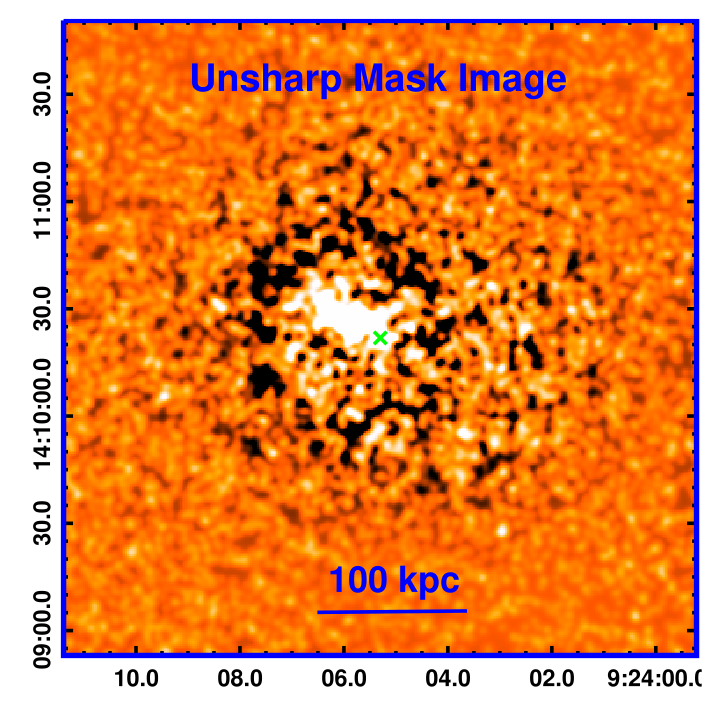}
\includegraphics[scale=0.46]{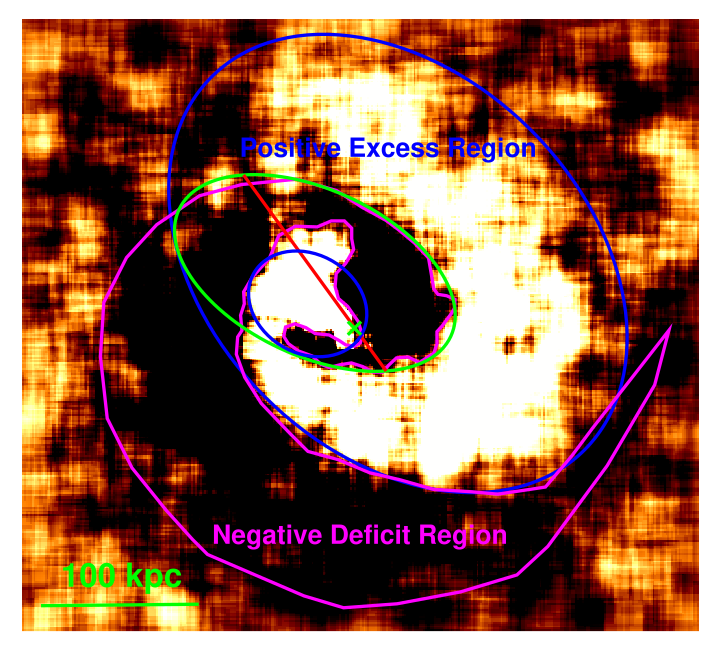}
}
\caption[Residual images] {\textit{Left panel:} 0.5-3 keV unsharp-mask image of A795 derived by subtracting 18 $\sigma$ wide Gaussian kernel smoothed image from that smoothed with a 2 $\sigma$ Gaussian kernel. Notice the structures in the ICM emission. Green cross represent the position of BCG. \textit{Right panel:} A central 3 $\times$ 3\arcmin\, 0.5-3 keV smooth model subtracted residual image of A795. This figure clearly reveals a spiral structure due to the sloshing in the ICM. The location of the BCG is indicated by a green cross.}
\label{resid}
\end{figure*}

\subsection{Radio data}
\label{radio_data_2.2}
The archival GMRT radio data at 325 MHz (Proposal Id: 17\_073) is used to study the radio properties of cluster A795. We also reanalysed the TGSS data at 150 MHz. Observation times on the cluster were 249 and 48 min for 325 and 150 MHz, respectively. Data was available with 32 MHz and 16 MHz bandwidth for 325 and 150 MHz, respectively, for RR and LL polarizations. Both the data sets were processed using the Source Peeling and Atmospheric Modeling (SPAM) pipeline. This pipeline includes all the standard data reduction steps, as well as the direction-dependent calibration to account for variations in visibility amplitude and phase caused by the antenna beam pattern and ionosphere. SPAM uses the bright sources within the primary beam for this purpose \citep[see][for more details]{intema_2009A&A...501.1185I, intema_2017A&A...598A..78I}. 3C286 was used by the SPAM to determine the flux scale and bandpass shape at both frequencies. The SPAM calibrated data sets were then imported to WCSCLEAN \citep{offringa_2014MNRAS.444..606O} for deriving images of different resolutions as described in Table \ref{tab:radio_imaging_parameter}. Primary beam correction was applied using the task wbpbgmrt\footnote{https://github.com/ruta-k/uGMRTprimarybeam} in CASA.

CASA viewer was used to measure the properties of radio sources. The flux densities (S) were measured within the 3$\sigma_{rms}$ contours and $\sigma_S = \sqrt{\sigma_c^2 + N_{beam}\sigma_{rms}^2}$ were used to calculate the error in the flux density measurements, where, $\sigma_c = 0.1S$ represent error due to calibration uncertainties and $N_{beam}$ is the number of beams across the diffuse emission.

\begin{table}
\caption{Parameters of the radio images} 
\label{tab:radio_imaging_parameter}  
\centering
\begin{tabular}{lccccr} 
\hline
Frequency &  Robust & u-v taper & FWHM & rms \\
(MHz)& (briggs) & ($\rm{arcsec}$) & (in \arcsec; PA in $^\circ$)& ($\mu$Jy/beam)\\
\hline  
150  &  uniform &  -- & $19.1\times17.5$; 40.4$^h$ & 3000\\
    & uniform & 15 &   $22.3\times19.9$; 50.0  &  2750     \\
    &  & inner uvcut $72\lambda$ &  &    \\
  &  0 &  20 & $29.8\times23.1$; 135.8$^l$  & 3600\\
325 & uniform & -- & $8.6\times6.8$; 67.6$^h$ & 210\\
    & uniform & 15 &   $18.7\times13.0$; 71.6    &  430     \\
        &  & inner uvcut $72\lambda$ &  &    \\
  &  0 & 12  & $20.7\times13.1$; 67.7$^l$& 360\\
\hline
\vspace{-0.3cm}
\end{tabular}
\begin{tablenotes}
      \small{
\item \textbf{Note}: Resolutions: h- high, l- low }
    \end{tablenotes}
\end{table}

\section{Results}
\label{results}
\subsection{X-ray image analysis}
\label{sec: x-ray image}
Exposure corrected, background subtracted 0.5-3 keV X-ray image of A795 (Fig.~\ref{raw}) revealed inhomogeneous structures in the ICM emission. To investigate the hidden features within the diffuse emission, we constructed a 0.5-3 keV unsharp-mask image and the residual map. An unsharp mask image was constructed by subtracting the X-ray image smoothed with a wider Gaussian kernel (18$\sigma$) from that smoothed with a narrower kernel (2$\sigma$). The resultant image is shown in Fig.~\ref{resid} (left panel) and depicts the spiral structure in the X-ray emission. For better visibility, we also derived its residual map. The residual map was prepared by subtracting a 2D smooth $\beta$-model of the X-ray emission from this cluster and was computed by fitting ellipses to the isophotes in the 0.5-3 keV cleaned, point source removed image of A795. The central 3\arcmin $\times$ 3\arcmin\, residual map of A795 is shown in Fig.~\ref{resid} (right panel) and for better presentation, it has been smoothed by a 10$\sigma$ wide Gaussian kernel. The residual map exhibits various structures in the ICM of this cluster. One of the interesting structures is the spiral-like excess emission (bright shades) extending from the South to the North in a counterclockwise direction on one side, while on the opposite side, it shows an X-ray deficit region (dark shades), fairly consistent with the results of \cite{2021MNRAS.503.4627U}. These spiral structures are found to cover a circular area with a radius of $\sim$80\arcsec (up to 186 kpc).

\subsection{Spectral Analysis} 
\label{SA}

The spectra were acquired using the CIAO script \textit{specextract} from locations exhibiting positive excess, indicated by blue ellipses that exclude the green portion, and negative deficit, indicated by magenta colour, within the energy range of 0.5-7 keV as shown in the right panel of Fig.~\ref{resid}. These spectra were then analyzed using standard models provided by \textsc{xspec v 12.9.1} \citep{1996ASPC..101...17A}. Normalized blank-sky background files were employed for background subtraction. The extracted counts after background subtraction were 16702 for the positive excess region and 7009 for the negative deficit region. Subsequently, the spectra were grouped so that each bin contained a minimum of 20 counts to enable the application of $\chi^2$ statistics. Each spectrum was fitted using a single temperature collisional equilibrium plasma model, \textsc{apec}, adjusted by the Galactic absorption column density ${\rm N_H= 2.89\times 10^{20}\, cm^{-2}}$ \citep{2005yCat.8076....0K}. The resulting spectral fits from the negative deficit region and positive excess region yielded ICM temperatures (kT) of $5.0\pm0.35$~keV and $4.1\pm0.15$~keV, respectively. The fits also provided metallicities (Z) of $0.33\pm0.06$ and $0.41\pm0.13$ Z$_{\odot}$ for the negative deficit and positive excess region, respectively. It is important to note that the reported errors in the X-ray spectral analysis correspond to the $90\%$ confidence level. The metallicities were determined relative to the metallicity value documented in \cite{1998SSRv...85..161G}. A comparison of temperature values suggests that the ICM within the spiral structure of the positive excess emission is comparatively cooler and we did not detect any considerable variation in the metallicity values in these spiral structures.

Furthermore, we tried to quantify the redshifts of the ICM from both these structures following \cite{2019ApJ...871..207U} to study the bulk motion of the ICM. hough the statistics involved in this investigation were limited, we roughly estimate the redshifts for both these regions. The total number of counts extracted from the positive excess and negative deficit regions are $\sim16702$ and $\sim7009$, respectively. We treated both these spectra independently with the same model as discussed earlier. Additionally, assuming the large space parameters to explore while fitting the spectra, we used the Monte Carlo Markov Chain (MCMC) tool. The chains were generated adopting the Goodman-Weare algorithm \citep{2010CAMCS...5...65G}, with 10 walkers, $10^4$\, burn-in steps and total length steps equal to $10^5$. The best-fit values of the redshift and the associated uncertainties for the positive excess and negative deficit regions are computed to be $0.1399^{+0.0062}_{-0.0051}$ and $0.147^{+0.0067}_{-0.0085}$, respectively. The estimated difference in the redshifts of the two regions is $\delta z$ = $0.0071^{+0.010}_{-0.008}$. With such large uncertainties associated with $\delta z$, it becomes challenging to make definitive statements about the bulk motion along the line of sight. Nevertheless, this redshift difference corresponds to an estimated velocity difference of $2128^{+2997}_{-2398}$ km~s$^{-1}$ for the ICM bulk motion.

\subsection{Diffuse radio emission in Abell 795}
\label{radio-diffuse}

\begin{figure}
\centering
\includegraphics[width=0.48\textwidth]{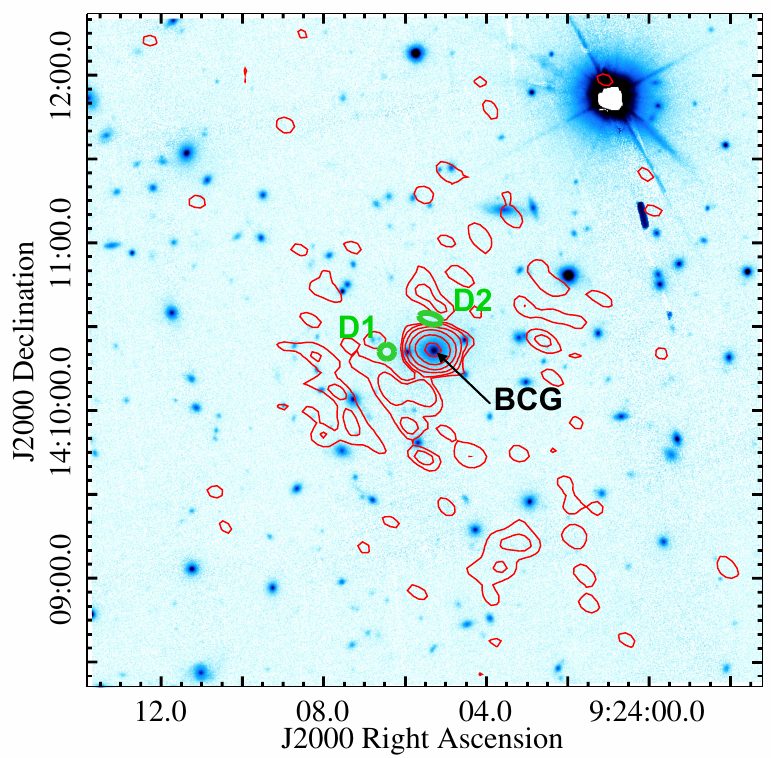}
\caption{Contours of high-resolution GMRT 325 MHz map at $5, 15, 45, ...\times \sigma= 210 \mu$Jy/beam are over-plotted on Pan-STARRS ‘r’ band optical image. Brightest Cluster Galaxy (BCG) and the two known X-ray cavities have been marked as BCG and D1 \& D2, respectively.}\label{fig: A795_high_opt}
\end{figure}

\begin{figure}
\centering
\includegraphics[width=0.48\textwidth]{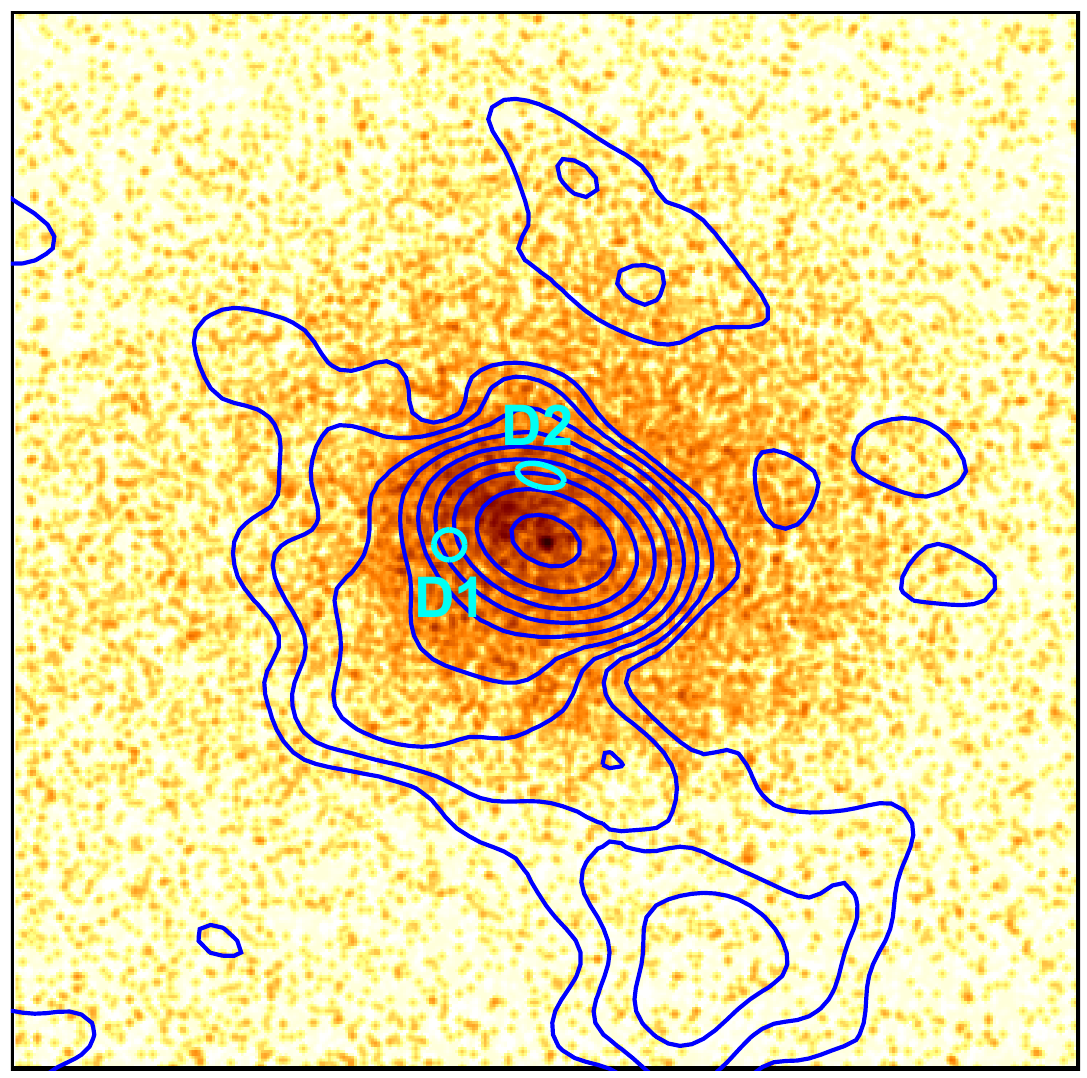}
\caption{ \textit{Chandra} X-ray map (Fig. \ref{raw}) over-plotted with the black contours of GMRT 325 MHz low-resolution map. Contours are plotted at $\pm 3, 6, 12, ... \sigma$ where $\sigma=360\;\mu$Jy/beam. D1 and D2 are two putative X-ray cavities as reported by \citealt{2021MNRAS.503.4627U}}\label{fig: A795_325_Xray}
\end{figure}

We detect an extended central diffuse radio emission in A795. Fig.~\ref{fig: A795_high_opt} shows the Pan-STARSS-1 ‘r’ band optical image of A795 with high-resolution radio contours at 325 MHz overlaid on it and exhibit morphology of the diffuse radio emission around the central BCG. \citet{2021MNRAS.503.4627U} have also reported this as a candidate radio mini-halo using various survey maps. However, the poor resolution and low sensitivity of the survey maps made it difficult to detect the actual extent and substructures of the diffuse emission in their study.

\begin{figure}
\centering
\includegraphics[width=0.48\textwidth]{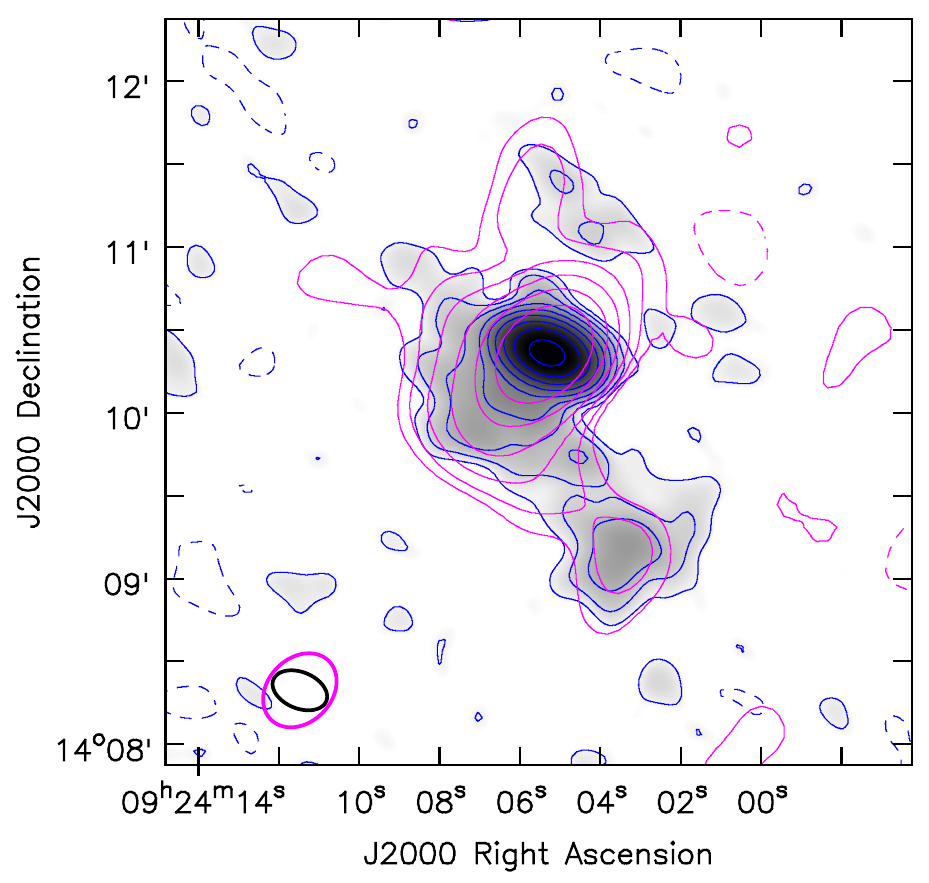}
\caption{Contours of 150 MHz map (magenta) and 325 MHz low-resolution map (blue) are superimposed on grey raster of 325 MHz low-resolution map at $\pm 3, 6, 12, ... \sigma$ where $\sigma=3.6\;$mJy/beam and $360\;\mu$Jy/beam for 150 and 325 MHz map, respectively}\label{fig: A795_325_150}
\end{figure}

In Fig.~\ref{fig: A795_325_Xray}, we displayed our low-resolution 325~MHz uGMRT contours overlaid on the \textit{Chandra} X-ray image (same as in Fig.~\ref{raw}). The details of all the radio maps used in this study are provided in Table~\ref{tab:radio_imaging_parameter}.
Fig.~\ref{fig: A795_325_150} provides a comparison of the radio maps at frequencies 150 and 325 MHz. The peak in central diffuse radio emission has been observed in both the GMRT maps and is found to coincide with the central BCG. This is surrounded by diffuse radio emission of average size $\sim75$\arcsec (180 kpc) evident in low-resolution maps at both the frequencies 150 and 325 MHz (see Fig. \ref{fig: A795_325_150}). A connected radio patch of about $70$\arcsec$\times 50$\arcsec ($170$~kpc $\times$~120~kpc) was also found towards the South-West at a distance of $\sim 75$\arcsec\, from the central peak. This patch and the extension towards north were reported as two small protrusions by \citet{2021MNRAS.503.4627U}. They speculated them to be remnants of old radio lobes, however, they could not confirm their nature due to the lack of multifrequency observations with adequate resolution and sensitivity in their study.

The radio emission associated with the central BCG was found to have flux density of $885 \pm 89$ mJy and $382 \pm 38$ mJy at uGMRT 150 and 325 MHz, respectively, as measured from our high-resolution maps (uniform weighted). Corresponding k-corrected radio power at 325 MHz was estimated to be $17.55\pm1.75\times10^{24}$ W Hz$^{-1}$. The radio patch seen on the southwest exhibits a flux density of $29.3\pm5.5$ mJy at 150 MHz and $22.1\pm2.5$ mJy at 325 MHz. Interestingly, the peak of this radio patch does not coincide with either of the optical sources in the field (see Fig.~\ref{fig: A795_high_opt} and \citealt{2021MNRAS.503.4627U}). Further, we subtracted the flux densities of the BCG (measured from high-resolution images) from the total flux densities of the entire diffuse radio emission (excluding the radio patch) measured from the low-resolution maps. The maps are created using uniform weighting, common inner uvcut of $72\lambda$ and uvtaper of $15\arcsec$ to report the flux densities corresponding to the large-scale diffuse radio emission associated with the ICM. The flux density of this diffuse source was found to be $365\pm38$ and $45.0\pm4.9$ mJy, respectively at 150 and 325 MHz. This corresponds to the k-corrected power of $2.54\pm0.28\times10^{24}$ W Hz$^{-1}$ at 325 MHz. All measured properties of the radio sources are summarised in Table \ref{tab:diffuse_emission}.

\begin{table*} 
\caption{Measured properties of the different components of the diffuse radio emission} %
\centering    
\small
\begin{tabular}{lccccc}
\hline       
Source &  \multicolumn{2}{c}{Flux density} & Spectral index & \multicolumn{2}{c}{Radio Power }\\
 &  \multicolumn{2}{c}{(mJy)} &  & \multicolumn{2}{c}{($10^{24}$ W Hz$^{-1}$)}\\
& [150MHz & 325 MHz] & $\alpha_{\rm{150}}^{\rm{325}}$ & [$P_{325MHz}$ & $P_{1.4GHz}$]\\
\hline  
BCG &  $885\pm89$ & $382\pm38$ & $-1.09 \pm 0.26$ & $17.55 \pm 1.75$ &$3.57 \pm 0.35$ \\
Diffuse emission & $365\pm38$ & $45.0\pm4.9$ & $-2.71 \pm 0.28$ & $2.54 \pm 0.28$  & $0.048\pm0.005$  \\
\hline
\end{tabular}
\label{tab:diffuse_emission} 
\end{table*}

\begin{figure}
{
\includegraphics[width=8.8cm]{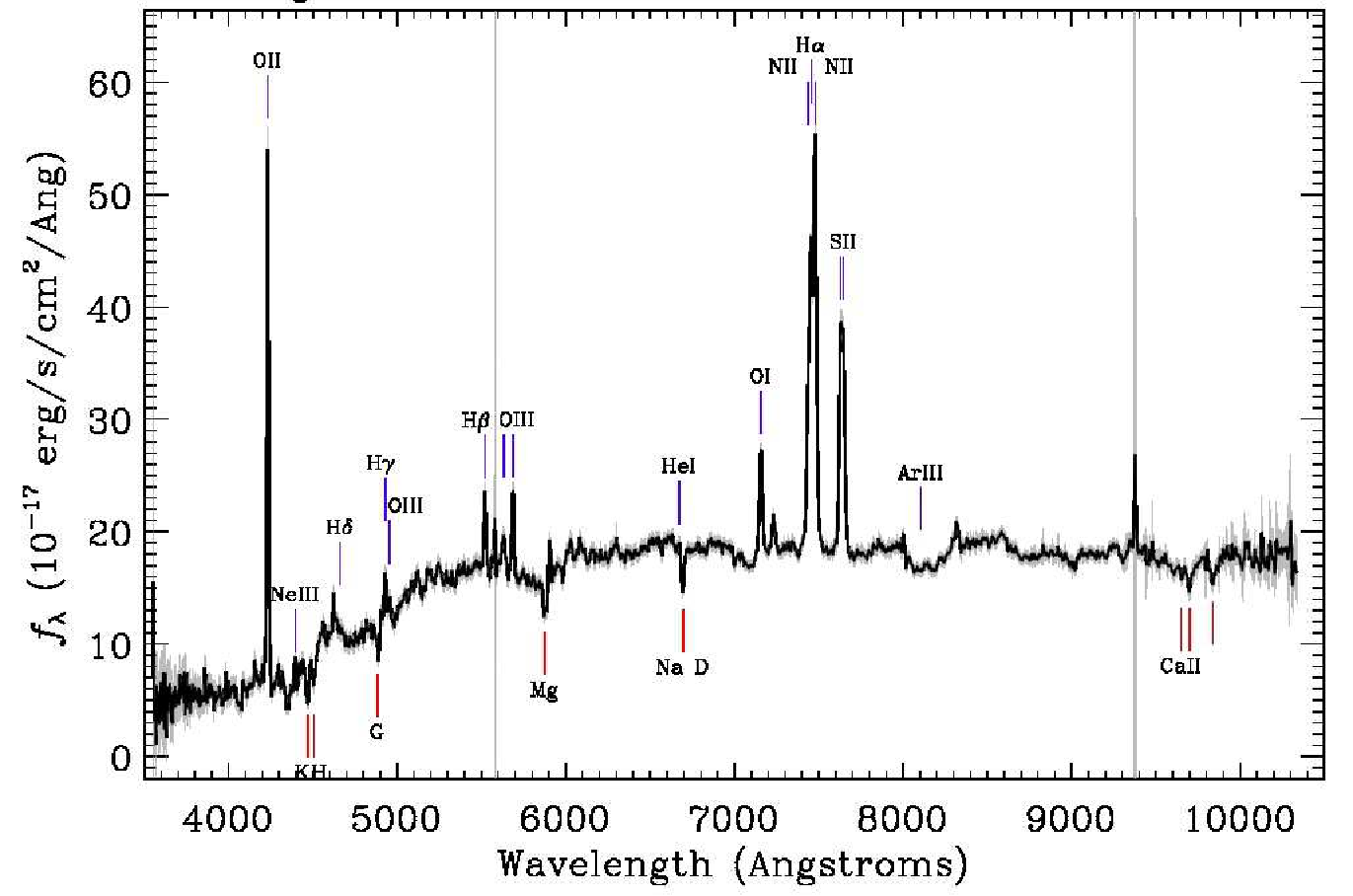}
}
\caption{Optical spectrum from SDSS of the BCG in A795 cluster.}
\label{sdss}
\end{figure} 

Using the flux densities obtained above, the integrated spectral index between the GMRT maps at 150 and 325 MHz was calculated for the radio sources. This analysis has revealed a steep spectral index ($\alpha_{BCG} = -1.09 \pm 0.26$) for the central BCG, while the extended diffuse emission around the BCG exhibits an ultra-steep spectral index ($\alpha_{MH} = -2.71 \pm 0.28$). The radio patch seen on the southwest shows a much flatter spectral index ($\sim -0.36$). The likelihood of the patch being the remnants of older activity, as discussed in \citet{2021MNRAS.503.4627U}, diminishes due to its flat spectrum. Moreover, we found no X-ray (possibly a missing cavity!) or optical counterparts for this emission. Consequently, the classification for this remains uncertain.

\citet{2021MNRAS.503.4627U} estimated a spectral index of $-0.93\pm0.09$ for the BCG using the data at higher frequencies (at 5 and 8.4 GHz). In the absence of a high-resolution image, the spectral index is then extrapolated to lower frequencies by the authors to estimate the contribution from BCG to the total diffuse emission. This possibly resulted in an underestimation of the radio flux for BCG at lower frequencies, leading to a flatter spectral index ($-1.73\pm0.86$) for the radio mini halo compared to what we have calculated ($\alpha_{MH}$ = $-2.71 \pm 0.28$). Though the spectral index values differ much, they are roughly consistent within the error bars.

\subsection{Star Formation Rate from optical spectral analysis}\label{sec:starform}

The spatial distribution of H$_\alpha$ emitting gas from Abell~795 exhibit quiescent simple elliptical and centrally concentrated morphology with an extent of $\sim11$~kpc \citep{Hamer_2016MNRAS}. It has been noticed by the authors that H$_\alpha$ flux decreased uniformly as the continuum decreased. The absence of prominent extended filaments of H$_\alpha$ is indicative of a stable system that has not experienced significant recent events to disturb the gas. Furthermore, the ratio of [NII] to H$_{\alpha}$ can provide insights into the excitation state of ionized gas and its sensitivity to AGN activity. Comparison of [NII]/H$_{\alpha}$ and [OI]/H$_{\alpha}$ in the sample (The sample was taken from ESO X-ray Cluster Elliptical Spectral Survey (EXCESS) which provided long slit spectra of 446 BCGs, further these spectra used to identify clusters with optical line emissions.) found majority of the objects are lie top right corner of plot (log[NII]/H$_{\alpha}$ $\ge$ -0.2 and log[OI]/H$_{\alpha}$ $\ge$ -0.9) suggest that ionisation is caused by non-stellar process such as AGN or in some cases AGN and shocks \citep{Hamer_2016MNRAS}. Hence, typically, AGN radiation fields are more harder than those produced by star formation \citep{Hamer_2016MNRAS}.Thus, the [NII] to H$_{\alpha}$ ratio tends to be highest in the central region of a BCG and in the areas where shock events may have occurred. In case of Abell~795, the [NII] / H$_{\alpha}$ ratio was found to be $1.5\pm0.85$ (log [NII]/H$_{\alpha}$=0.18) \citep{Hamer_2016MNRAS}. This indicates that the gas in the central region is excited by more energetic source, such as AGN by the photoionization method.

\begin{figure}
{
\includegraphics[scale=0.47]{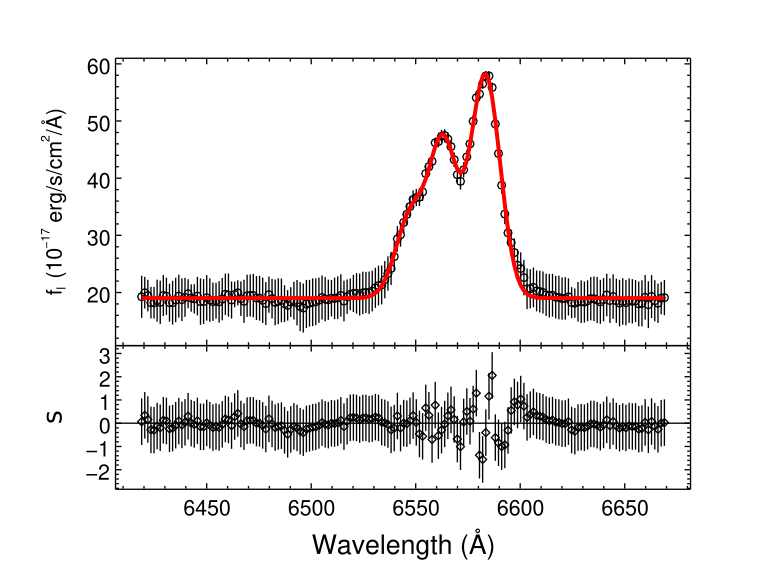}
}
\caption{Cosmological redshift corrected BCG spectrum of Abell~795 fitted with three Gaussian 1-D and continuum models.}
\label{gauss}
\end{figure}

Indeed, cluster A795 is found to host a pair of putative X-ray cavities indicative of radio jet feedback from the central AGN \citep{2021MNRAS.503.4627U}. The excess X-ray emission in the central region of this cluster is apparent in Fig.~\ref{resid} suggesting that a large fraction of the cool ICM is being deposited in the central region. To investigate whether this deposited ICM yields star formation in the BCG, we made use of the H$_{\alpha}$ emission flux densities as recorded in Sloan Digital Sky Survey (SDSS) DR13 \citep{2013AJ....145...10D}. 
Fig.~\ref{sdss} clearly depicts strong optical emission lines of the BCG in A795 cluster. The H$_{\alpha}$ flux density was computed by fitting Gaussian profile to it. Fig.~\ref{gauss} shows the best-fit spectrum, which was obtained by fitting three Gaussian components along with the continuum. The best fit model yielded the H$_{\alpha}$ emission peak at 6563.18$^{+1.18}_{-0.85}$ \AA\,with the FWHM=15.73$^{+2.52}_{-2.20}$ \AA\,. In addition to the H$_{\alpha}$ emission line, we also fitted both the [NII] lines positioned at 6547.69 \AA\ and 6583.34 \AA. The best fit parameters are listed in Table~\ref{ele} and yielded the continuum subtracted H$_{\alpha}$ flux density equal to $4.19\,\times\,10^{-15}\,$ erg\,  s$^{-1}$~cm$^{-2}$, corresponding to $L_{H_{\alpha}} = 20.78\, \times\, 10^{40}\, $ erg\,  s$^{-1}$.

\begin{figure}
\centering
\includegraphics[width=0.49\textwidth]{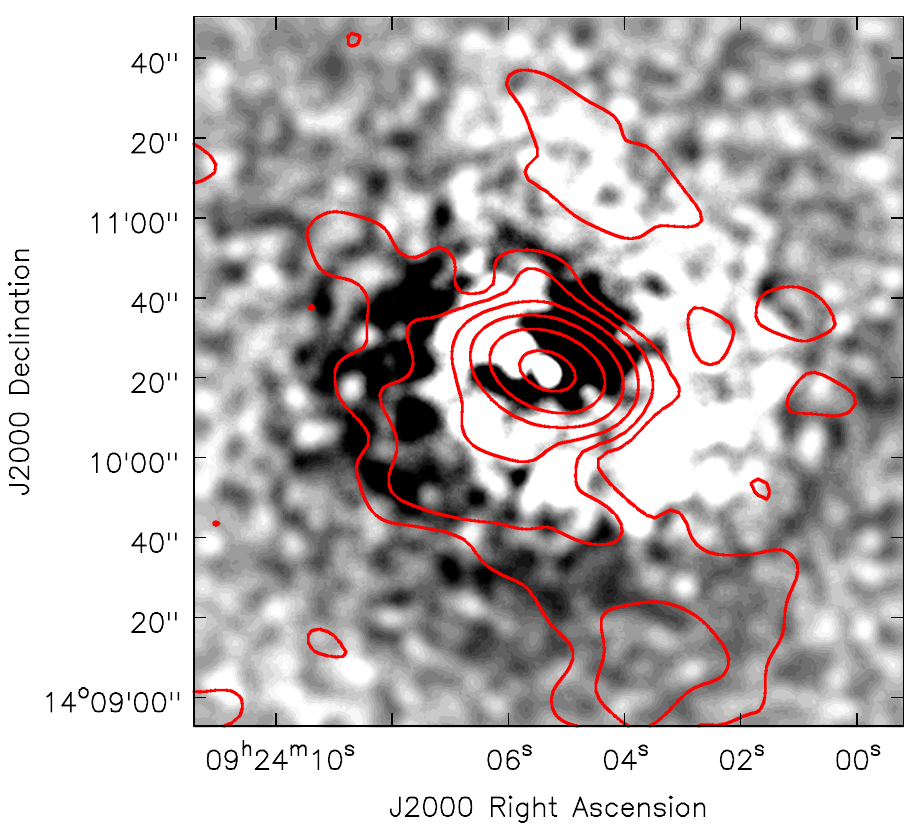}
\caption{Residual X-ray image (right panel of Fig.~\ref{resid}) of A795 over-plotted with contours of 325 MHz low-resolution image at $3, 9, 27, .. \times 360 \mu$Jy/beam.}
\label{fig: A795_spiral_MH}
\end{figure}

\begin{table*}
 \caption{The best-fit parameters of the Gaussian model fitted to the emission lines. All the reported error bars are at 1-sigma level. The best-fit statistic was found to be $\chi^2/dof=161/152$.}
    \centering
    \begin{tabular}{|ccccc|} \hline 
     Element & model   &   Position	  &  FWHM      &   Amplitude    \\
		--	 &   --    &  (\AA)		  & (\AA)      &  $10^{-17}\,\mathrm{erg\,s^{-1}\,cm^{-2}\,A^{-1}}$   \\
\hline 
   H$_{\alpha}$ & Gaussian 1-D &  6563.18$^{+1.18}_{-0.85}$   &  15.7287$^{+2.52}_{-2.20}$  &  26.6033$^{+1.50}_{-6.51}$ \\ \\
 \big[NII\big]	 &	Gaussian 1-D & 6583.34$^{+0.28}_{-0.24}$   &  16.6074$^{+0.65}_{-0.66}$  &  38.8849$^{+0.39}_{-0.42}$ \\ \\
\big[NII\big]   & Gaussian 1-D &  6547.69$^{+3.89}_{-2.01}$   &  16.7665$^{+6.24}_{-3.44}$  &  14.2479$^{+2.56}_{-2.23}$ \\ \\
Continuum & Constant &  --	&  --  & 19.1 \\
    \hline
    \end{tabular}
    \label{ele}
\end{table*}

We also used the FUV flux densities as measured by the GALEX-MAST FUV imager. This resulted in to a net background-subtracted source count rate of 0.101 from within 10$\arcsec$ circular regions (visual inspection method), which was then converted to the resultant flux equal to $2.18\,\times\,10^{-13}\,$ erg\,  s$^{-1}$~cm$^{-2}$. The equivalent FUV luminosity was found to be equal to $L_{FUV} = 10.87\, \times\, 10^{42}\,$ erg~ s$^{-1}$. 
Both the $H_{\alpha}$ and FUV luminosities thus estimated were then used to compute the star formation rate (SFR) in Abell~795 using the equations from \cite{2006ApJS..164...38I,1998ARA&A..36..189K}, respectively, for the FUV and $H_{\alpha}$ luminosities. 

\begin{equation*}
\frac{SFR(FUV)}{M_{\odot} yr^{-1}} = 2.61 \times 10^{-43.51}\, L_{FUV}(\rm{erg~s^{-1}}) 
\end{equation*}
\begin{equation*}
\frac{SFR(H_{\alpha})}{M_{\odot} yr^{-1}} =  7.9 \times 10^{-42}\, L_{H_{\alpha}}(\rm{erg\, s^{-1}})   
\end{equation*}
The resultant star formation rates from the FUV and the $H_{\alpha}$ flux densities were found to be $\sim$ 0.88 $M_{\odot} yr^{-1}$ and $\sim$ 1.64 $M_{\odot} yr^{-1}$, respectively.

\section{Discussion}\label{disc}

The multi-waveband analysis revealed complex energy distribution in this cool-core cluster A795. Though the cool core and H$_\alpha$ morphologies point towards a relaxed state, the presence of substructures, such as spiral features, is clearly evident in the \chandra\ X-ray residual map as well as in the unsharp-mask image. The origin of such spiral structures in X-rays is usually the imprint of the sloshing of gas in the cluster due to recent minor mergers or other dynamical activities. While this spiral feature in A795 has earlier been reported in \citet{2021MNRAS.503.4627U}, additionally, we report an excess emission on one side and an X-ray deficit counterpart of it on the other side of the spiral within 186 kpc region of the cluster core (Fig.~\ref{resid} right panel). These log spiral structures apparent in the positive excess emission (bright shade) and the negative deficit emission (dark shade) are complimenting each other and have almost similar extents. The multi-phase ICM, including radio-emitting gas (as found in our GMRT maps) within the central 186 kpc collectively point towards the fact that the ICM in this cluster is not fully in hydrostatic equilibrium as may be expected due to its cool-core features.

X-ray spectral analysis of the ICM extracted independently from both these regions revealed consistently contrasting properties. However, this is not surprising as many similar cases were reported earlier \cite[e.g.][]{2004ApJ...616..178C,2011ApJ...737...99B,2013ApJ...773..114P,2014A&A...570A.117G,2017ApJ...837...34U}. Importantly, though, the positive excess emission region (brighter shade) in A795 appears to be cooler relative to that from the negative deficit emission region (darker shade). Also, we did not detect any appreciable variation in the metallicity values in the spirals, like the case of Abell 1835 \citep{2017ApJ...837...34U}. As the ICM in this cluster is revealed to be disturbed by the AGN feedback as well as minor mergers, such an environment does not promote star formation in the central BCGs \citep{2007ARA&A..45..117M,2012NJPh...14e5023M}. Our estimated star formation in the BCG, employing the SDSS's $H_{\alpha}$ and the GALEX's FUV flux density measurements are at $\sim$1.64 and $\sim$0.88 M$_{\odot}$yr$^{-1}$, respectively. Here, it is likely that the H$_{\alpha}$ and the UV emissions probe different epochs of the star formation, therefore, the collective star formation at most could be 2.52~M$_{\odot}$~yr$^{-1}$. This estimated SFR in this cluster is significantly lower than the reported mass deposition rate of 14~M$_{\odot}$~yr$^{-1}$ \citep{Ubertosi_2021} possibly due to star formation quenching from AGN feedback, reminiscent of observations in other cool-core clusters \citep{2006MNRAS.366..417F,2009ApJ...697..867K,2012MNRAS.421..808P,2019MNRAS.485.1981V,2019MNRAS.484.4113K}.

In addition to the X-ray analysis of the cluster core, here, we present the first deep radio study of cluster A795 at GMRT 325MHz. Our comparatively high resolution radio images at GMRT 150 and 325 MHz revealed a previously indicated extended diffuse radio structure \citep{2021MNRAS.503.4627U} of size $\sim180$~kpc around the central BCG allowing us to compute the spectral properties more reliably. The radio emission is found to be ultra-steep ($\alpha_{150\rm{MHz}}^{\rm{325MHz}}=-2.71$) at low frequencies, covering almost the full extent of the X-ray cool-core region (see Fig. \ref{fig: A795_spiral_MH}). It is also confirmed that the extension of the radio emission is well confined within the spiral pattern and two cold fronts that are located at the outskirts of these spirals. Moreover, a weak cool-core in addition to the spirals and cold fronts suggests large-scale sloshing in the cluster core, overall a suitable environment for harbouring radio mini-haloes \citep{gitti_2002A&A...386..456G, zuhone_2013ApJ...762...78Z}. Notably, while the radio emission covers the entire dark shaded spiral arm following the region of X-ray deficit emission (see Fig.~\ref{fig: A795_spiral_MH}), there is a clear lack of radio emission where the bright shaded spiral arm is located. This is therefore supporting the fact that X-ray deficiency in clusters is created due to the replacement of thermal gas by non-thermal \cite[reviewed in][]{2007ARA&A..45..117M, 2012ARA&A..50..455F}. Usually, such X-ray depreciation is expected to be caused by the inflation of radio lobes. Intriguingly, in the present case, the X-ray deficit is seemingly filled by mini-halo-like radio emission and not directly by the AGN feedback. Furthermore, \citet{2021AN....342.1207U} also reported two putative X-ray cavities approximately marked as D1 and D2 in Fig.~\ref{fig: A795_325_Xray}. But, our high-resolution 325~MHz GMRT map (see Fig.~\ref{fig: A795_high_opt}) clearly shows no physical connection with AGN feedback to those cavities. Rather, the extended and diffuse radio emission seen in Fig.~\ref{fig: A795_325_Xray} indicates a link with the mini-halo-type emission to these cavities.

\begin{figure}
{
\includegraphics[width=0.48\textwidth]{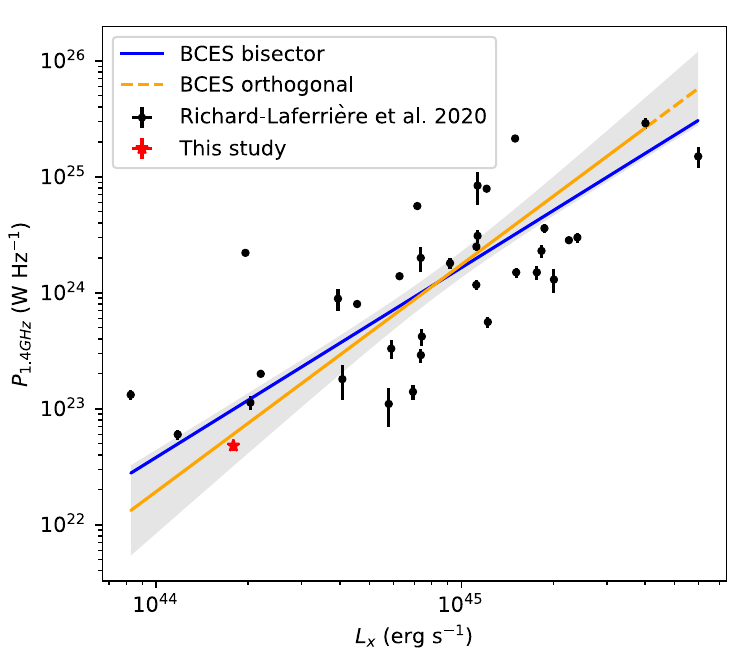}
}
\caption{Radio mini halo power ($P_{1.4GHz}$) and the X-ray luminosity within a radius of 600 kpc ($L_x$) plot (reproduced from \citealt{richard_2020MNRAS.499.2934R}). The graph displays both the best-fitting lines obtained from the BCES-orthogonal method (orange) and the BCES-bisector method (blue). Furthermore, the gray region on the graph represents the 95\% confidence regions of the best-fitting relation obtained from the BCES-orthogonal method. The red point represents the cluster in study for which radio power at 1.4 GHz is calculated using the spectral index ($\alpha_{150MHz}^{325MHz} = -2.71\pm0.28$) given in Table \ref{tab:diffuse_emission}.}
\label{fig: correlations}
\end{figure}

Nevertheless, comprehending the origin of such mini-halo-type diffuse radio emission in this cluster is a daunting task given the fact that it requires answers to two significant inquiries. The first is to understand the original source of particle injection that finally emits through the synchrotron process, while, the second concerns the particle acceleration mechanism that makes them capable of emitting synchrotron radiations. Distinctly determining the source of the electron population, i.e., whether it is associated with the thermal component of the cluster ICM or the brightest cluster galaxy (BCG), especially in cool-core clusters with BCG remains a challenge. In this context, the radio structure in this cluster appears to be asymmetric around the BCG and the spectrum is significantly steeper than the usual BCG spectrum. Additionally, the measured physical parameters of A795 were found to follow the correlation between the radio mini halo power and core-excised X-ray luminosity (see Fig.~\ref{fig: correlations}). All these indicate that the diffuse radio emission here is possibly linked to the ICM and not directly to the central BCG. Nevertheless, the dynamical state, confinement of the extended diffuse radio emission around the cool-core by the cold fronts of the cluster and the strong correlation of radio power ($P_{1.4GHz}$) with the X-ray luminosity ($L_{x}$) etc. support the mini-halo nature of the observed diffuse radio emission. However, there is an inherent uncertainty associated with this classification, especially when taking into account the ultra-steep spectral index and asymmetric morphology. Notably, the occurrence of such asymmetric morphologies in mini-haloes has been recently observed in certain clusters (e.g., MS 1455.0+2232, \citealt{riseley_2022MNRAS.512.4210R}), a phenomenon also discussed in this paper.

Even though the cool core in a cluster is a strong sign of relaxed nature, almost all other observed features in A795 were found to be related to its dynamical evolution of multi-phase ICM. The quantification of dynamical motions of different regions in the ICM of A795 was done on the basis of spatially resolved X-ray redshift measurements. The observed redshift distribution also reveals that the ICM has a sloshing motion most probably due to multiple sources of perturbation. The measured value of velocity difference (see Sec.~\ref{SA}) between these two regions is a sign of gas sloshing capable enough of inducing turbulence in the ICM and may lead to the formation of the observed spiral structure and for that instance the radio mini-halo. The simulation by \cite{Keshet_2012} provides a possible theoretical explanation for the generation of spirals in cool-core clusters. Where, in a modelled (quasi-steady-state) multi-phase system with a fast outflowing cold gas and a hot and slow inflow phase found to produce a prominent spiral at the core. The outflow in the system is regulated by some low level of mass accretion, similar to that of an AGN outflow and an accretion-driven feedback loop. We found more support for this two-phase scenario with differential velocity, which not only produces spirals but also sufficiently quenches the star formation due to weak cooling flows and facilitates the smooth metallicity distribution by mixing the gas on large scales, perfectly in tune with the observed low star formation rate and no appreciable variation in the metal distribution in A795. The AGN outflows (i.e., feedback) seem to interact with the spiralling flow, and possibly entrained in the motion and get distributed in the ICM removing the hot X-ray emitting gas (See Fig.~7 in \citealt{Keshet_2012}). This is what we also see in A795, where one arm of the spiral, which is X-ray deficient, is filled with non-thermal diffuse radio emission. Thus we may conjecture that the low-level turbulence generated due to sloshing in A795 has (re-)accelerated the electrons diffused in the ICM from the central BCG and produced the observed radio mini-halo-type emission \citep{pformmer_2004A&A...413...17P} having the shape similar to the X-ray deficient part of the X-ray spiral.

\section{Summary and conclusion}
\label{con}

We present the results from our analysis of X-ray and radio data of cluster Abell~795. We re-analysed high-resolution 30~ks \textit{Chandra} data and complemented it with the newly analysed GMRT 325 and 150~MHz data, the SDSS~DR13 optical spectra as well as GALEX’s FUV data. Our multi-waveband information revealed a complex multi-phase ICM in this previously reported cool-core cluster. The object shows a mixed characteristic of a relaxed cluster having X-ray substructures in the form of a log spiral structure of positive excess emission on one side and its X-ray deficit counterpart on the other side. Our X-ray redshift measurements for these two spiral counterparts hint at a velocity difference between them. This could potentially have triggered large-scale gas sloshing in the core of the cluster. This finding helped us to argue in favour of a two-phase scenario with a differential velocity that in the simulation setup was found to produce similar sloshing spirals. It was shown to drive turbulence in the cool-core region that would cause the re-acceleration of ICM cosmic electrons or secondary electrons to produce radio mini-halo-type diffuse emissions. The set-up has also favoured the observed quenched star formation rate of 2.52~M$_{\odot}$~yr$^{-1}$ (total from FUV and H$_{\alpha}$) in the central BCG of A795.

Interestingly, our GMRT 150~\&~325 MHz radio maps unravelled an unusually steep-spectrum diffuse radio source of almost the same extent ($\sim180$~kpc) as the spirals ($\sim186$~kpc) placed around the central BCG and enclosed within two antipodal cold fronts. The diffuse radio emission in the cluster has been putatively found to be a radio mini-halo owing to its co-spatial existence with the cool core and the association with the sloshing features. A further peculiar feature is, while this diffuse radio source well covers the X-ray deficit spiral structure - possibly by replacing the thermal gas in this region, it is almost missing on the counter log spiral region of X-ray excess.

\section*{Acknowledgments} 
{SKK gratefully acknowledges the financial support of UGC, New Delhi under the Rajiv Gandhi National Fellowship (RGNF) Program. SP wants to thank the DST INSPIRE faculty scheme
(code: IF-12/PH-44) for funding his research group. SS acknowledges the Department of Atomic Energy for funding support, under project 12-R\&D-TFR-5.02-0700. NDV and MKP gratefully acknowledge IUCAA, Pune for providing the E-Library facility. We sincerely thank the anonymous referee for her/his insightful suggestions. The data for this work has been obtained from the \chandra\, Data Archive, NASA/IPAC Extragalactic Database (NED), High Energy Astrophysics Science Archive Research Center (HEASARC), Sloan Digital Sky Survey (SDSS) and The Galaxy Evolution Explorer (GALEX), PanSTARRS-1 imaging data. This work has made use of software packages CIAO and Sherpa provided by the \chandra\, X-ray Center.
}

\section*{DATA AVAILABILITY}
The raw data underlying this article are publicly available in the {\it Chandra} (https://cda.harvard.edu/chaser/) and GMRT archives (https://naps.ncra.tifr.res.in). Analysed data may be made available at a reasonable request to the communicating authors.

\bibliographystyle{mnras}
\bibliography{mybib_a795,mybib} 

\bsp	
\label{lastpage}
\end{document}